\documentclass[journal]{IEEEtran}
\usepackage{graphicx}

\usepackage{cite}

\ifCLASSINFOpdf
\else
\fi
\begin{document}
%
\title{Artificial Intelligence\\ for Satellite Communication: A Review}
%
%
%

\author{
        Fares~Fourati,
        Mohamed-Slim~Alouini,~\IEEEmembership{Fellow,~IEEE}
\thanks{Fares Fourati and Mohamed Slim Alouini are with King Abdullah University of Science and Technology (KAUST), CEMSE Division, Thuwal, 23955-6900 KSA, (e-mail: fares.fourati@kaust.edu.sa, slim.alouini@kaust.edu.sa)}
}

%
%

\markboth{JAN. 2021}
{\MakeLowercase{\textit{et al.}}: Artificial Intelligence for Satellite Communication}
%



\maketitle 

\begin{abstract}

Satellite communication offers the prospect of service continuity over uncovered and under-covered areas, service ubiquity, and service scalability. However, several challenges must first be addressed to realize these benefits, as the resource management, network control, network security, spectrum management, and energy usage of satellite networks are more challenging than that of terrestrial networks. Meanwhile, artificial intelligence (AI), including machine learning, deep learning, and reinforcement learning, has been steadily growing as a research field and has shown successful results in diverse applications, including wireless communication. In particular, the application of AI to a wide variety of satellite communication aspects have demonstrated excellent potential, including beam-hopping, anti-jamming, network traffic forecasting, channel modeling, telemetry mining, ionospheric scintillation detecting, interference managing, remote sensing, behavior modeling, space-air-ground integrating, and energy managing. This work thus provides a general overview of AI, its diverse sub-fields, and its state-of-the-art algorithms. Several challenges facing diverse aspects of satellite communication systems are then discussed, and their proposed and potential AI-based solutions are presented. Finally, an outlook of field is drawn, and future steps are suggested.
\end{abstract}


%
\IEEEpeerreviewmaketitle

\newcommand{\RNum}[1]{\uppercase\expandafter{\romannumeral #1\relax}}

\section{Introduction}
\IEEEPARstart{T}{he} remarkable advancement of wireless communication systems, quickly increasing demand for new services in various fields, and rapid development of intelligent devices have led to a growing demand for satellite communication systems to complement conventional terrestrial networks to give access over uncovered and under-covered urban, rural, and mountainous areas, as well as the seas.

There are three major types of satellites, including the geostationary Earth orbit,  also referred to as a geosynchronous equatorial orbit (GEO), medium Earth orbit (MEO), and low Earth orbit (LEO) satellites. This classification depends on three main features, i.e., the altitude, beam footprint size, and orbit. GEO, MEO, and LEO satellites have an orbit around the Earth at an altitude of 35786 km, 7000–25000 km, and 300–1500 km, respectively. The beam footprint of a GEO satellite ranges from 200 to 3500 km; that of an MEO or LEO beam footprint satellite ranges from 100 to 1000 km. The orbital period of a GEO satellite is equal to that of the Earth period, which makes it appear fixed to the ground observers, whereas LEO and MEO satellites have a shorter period, many LEO and MEO satellites are required to offer continuous global coverage. For example, Iridium NEXT has 66 LEO satellites and 6 spares, Starlink by SpaceX plans to have 4425 LEO satellites plus some spares, and O3b has 20 MEO satellites including 3 on-orbit spares \cite{book1}.

Satellite communication use cases can also be split into three categories: i) service continuity, to provide network access over uncovered and under-covered areas; ii) service ubiquity, to ameliorate the network availability in cases of temporary outage or destruction of a ground network due to disasters; and iii) service scalability, to offload traffic from the ground networks. In addition, satellite communication systems could provide coverage to various fields, such as the transportation, energy, agriculture, business, and public safety fields \cite{intro11}.

Although satellite communication offers improved global coverage and increased communication quality, it has several challenges. Satellites, especially LEO satellites, have limited on-board resources and move quickly, bringing high dynamics to the network access. The high mobility of the space segments, and the inherent heterogeneity between the satellite layers (GEO, MEO, LEO), the aerial layers (unmanned aerial vehicles (UAVs), balloons, airships), and the ground layer make network control, network security, and spectrum management challenging. In addition, achieving high energy efficiency for satellite communication is more
challenging than for terrestrial networks.

Several surveys have discussed different aspects of satellite communication systems, such as handoff schemes \cite{intro1}, mobile satellite systems \cite{intro2}, MIMO over satellite \cite{intro3}, satellites for the Internet of Remote Things \cite{intro4}, inter-satellite communication systems \cite{intro5},  Quality of Service (QoS) provisioning \cite{intro6}, space optical communication \cite{intro7}, space-air-ground integrated networks \cite{sagin1}, small satellite communication \cite{intro9}, physical space security \cite{intro10}, CubeSat communications \cite{intro12}, and non-terrestrial networks \cite{intro11}. 
Meanwhile, interest in artificial intelligence (AI) increased in recent years. AI, including machine learning (ML), deep learning (DL) and reinforcement learning (RL), has shown successful results in diverse applications in science and engineering fields, such as electrical engineering, software engineering, bioengineering, financial engineering, and medicine etc. Several researchers have thus turned to AI techniques to solve various challenges in their respective fields and have designed diverse successful AI-based applications, to overcome several challenges in the wireless communication field.

Many researchers have discussed AI and its applications to wireless communication in general \cite{w0,w1,w2,w3}. Others have focused on the application of AI to one aspect of wireless communication, such as wireless communications in the Internet of Things (IoT) \cite{iot}, network management \cite{mng}, wireless security \cite{sec}, emerging robotics communication \cite{rob}, antenna design \cite{reflect1} and UAV networks \cite{uav1,uav2}. Vazquez et al. \cite{sat1} briefly discussed some promising use cases of AI for satellite communication, whereas Kato et al. \cite{sagin2} discussed the use of AI for space-air-integrated networks. The use of DL in space applications has also been addressed \cite{sat3}.

\begin{figure}
    \centering
    \includegraphics[scale = 0.53]{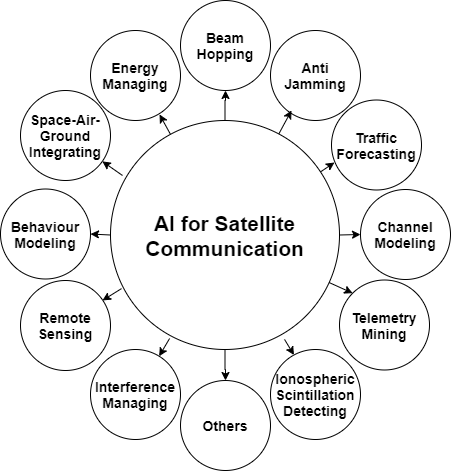}
    \caption{Applications of artificial intelligence (AI) for different satellite communication aspects}
    \label{apps}
\end{figure} 

\begin{table}
\setlength{\tabcolsep}{3pt}
\begin{tabular}{|p{35pt}|p{180pt}|}
\hline
AE & Autoencoder\\
\hline
AI & Artificial intelligence\\
\hline
AJ & Anti-jamming\\
\hline
ARIMA & Auto regressive integrated moving average \\
\hline
ARMA & Auto regressive moving average \\
\hline
BH & Beam hopping\\
\hline
CNN & Convolutional neural network \\
\hline
DL & Deep learning \\
\hline
DNN & Deep neural network \\
\hline
DRL & Deep reinforcement learning \\
\hline
ELM & Extreme learning machine\\
\hline
EMD & Empirical mode decomposition \\
\hline
FARIMA & Fractional auto regressive integrated moving average \\
\hline
FCN & Fully convolutional network\\
\hline
FDMA & Frequency division multiple access\\
\hline
FH & Frequency hopping\\
\hline
GA & Genetic algorithms\\
\hline
GANs & Generative adversarial networks\\
\hline
GNSS & Global navigation satellite system \\
\hline
IoS & Internet of satellites\\
\hline
kNN & k-nearest neighbor \\
\hline
LRD & Long-range-dependence \\
\hline
LSTM & Long short-term memory \\
\hline
MDP & Markov decision process \\
\hline
ML & Machine learning\\
\hline
MO-DRL & Multi-objective deep reinforcement learning\\
\hline
NNs & Neural networks \\
\hline
PCA & Principal component analysis\\
\hline
QoS & Quality of service\\
\hline
RFs & Random forests\\
\hline
RL & Reinforcement learning\\
\hline
RNNs & Recurrent neural networks\\
\hline
RS & Remote sensing\\
\hline
RSRP & Reference signal received power\\
\hline
SAGIN & Space-air-ground integrated network\\
\hline
SRD & Short range dependence\\
\hline
SVM & Support vector machine\\
\hline
SVR & Support vector regression\\
\hline
SatIot & Satellite Internet of Things\\
\hline
UE & User equipment \\
\hline
VAEs & Variational autoencoders \\
\hline
\end{tabular}
\label{table}
\caption{Acronyms and Abbreviations}
\end{table}

Overall, several researchers have discussed wireless and satellite communication systems, and some of these have discussed the use of AI for one or a few aspects of satellite communication; however, an extensive survey of AI applications in diverse aspects of satellite communication has yet to be performed.

This work therefore aims to provide an introduction to AI, a discussion of various challenges being faced by satellite communication and an extensive survey of potential AI-based applications to overcome these challenges. A general overview of AI, its diverse sub-fields and its state-of-the-art algorithms are presented in Section \RNum{2}. Several challenges being faced by diverse aspects of satellite communication systems and potential AI-based solutions are then discussed in Section \RNum{3}; these applications are summarized in Fig.\ref{apps}. For ease of reference, the acronyms and abbreviations used in this paper are presented in Table \ref{table}.

\section{Artificial Intelligence (AI)}
The demonstration of successful applications of AI in healthcare, finance, business, industries, robotics, autonomous cars and wireless communication including satellites has led it to become a subject of high interest in the research community, industries, and media. 

This section therefore aims to provide a brief overview of the world of AI, ML, DL and RL. Sub-fields, commonly used
algorithms, challenges, achievements, and outlooks are also addressed.

\subsection{Artificial Intelligence}
\begin{figure}
    \centering
    \includegraphics[scale = 0.36]{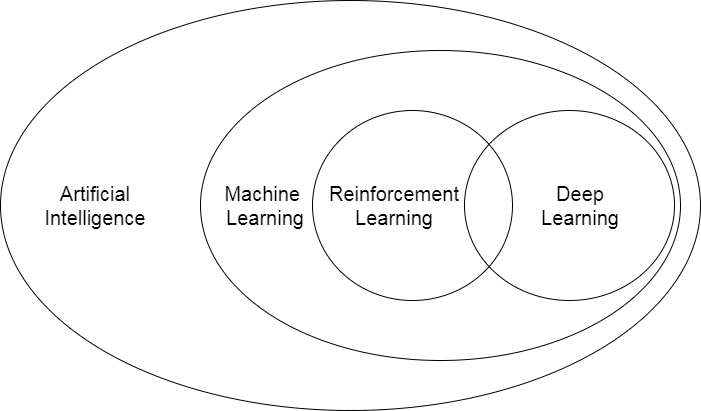}
    \caption{Artificial Intelligence, Machine Learning, Deep Learning and Reinforcement Learning}
    \label{AIMLDLRL}
\end{figure}

Although AI sounds like a novel approach, it can be traced to the 1950s and encompasses several approaches and paradigms. ML, DL, RL and their intersections are all parts of AI, as summarized in Fig.\ref{AIMLDLRL} \cite{b1}. Thus, a major part of AI follows the learning approach, although approaches without any learning aspects are also included. Overall, research into AI aims to make the machine smarter, either by following some rules or by facilitating guided learning. The former refers to symbolic AI; the latter refers to ML. Here smarter indicates the ability to accomplish complex intellectual tasks normally necessitating a human such as classification, regression, clustering, detection, recognition, segmentation, planning, scheduling, or decision making. In the early days of AI, many believed that these tasks could be achieved by transferring human knowledge to computers by providing an extensive set of rules that encompasses the humans' expertise. Much focus was thus placed on feature engineering and implementing sophisticated handcrafted commands to be explicitly used by the computers. Although this symbolic AI has been suitable for many applications, it has shown various limitations in terms of both precision and accuracy for more advanced problems that show more complexity, less structure, and more hidden features such as computer-vision and language-processing tasks. To address these limitations, researchers turned to a learning approach known as ML.

\subsection{Machine Learning (ML)}

\begin{figure}
    \centering
    \includegraphics[scale = 0.36]{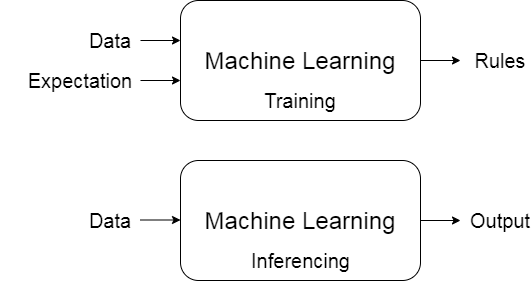}
    \caption{Machine Learning Approach}
    \label{ML}
\end{figure} 

ML, which encompasses DL and RL, is a subset of AI. In contrast to symbolic AI, where the machine is provided with all the rules to solve a certain problem, ML requires a learning approach. Thus, rather than giving the rules to solve a problem, the machine is provided with the context to learn the rules by itself to solve the issue, as shown in Fig.\ref{ML} and best summarized by the AI pioneer Alan Turing \cite{b2}: "An important feature of a learning machine is that its teacher will often be very largely ignorant of quite what is going on inside, although he may still be able to some extent to predict his pupil's behavior,'' An ML system is trained rather than programmed with explicit rules. The learning process requires data to extract patterns and hidden structures; the focus is on finding optimal representations of the data to get closer to the expected result by searching within a predefined space of possibilities using guidance from a feedback signal, where representations of the data refer to different ways to look at or encode the data. To achieve that, three things are mandatory: input data, samples of the expected output, and a way to measure the performance of the algorithm \cite{b1}. This simple idea of learning a useful representation of data has been useful in multiple applications from image classification to satellite communication. 

\begin{figure}
    \centering
    \includegraphics[scale = 0.36]{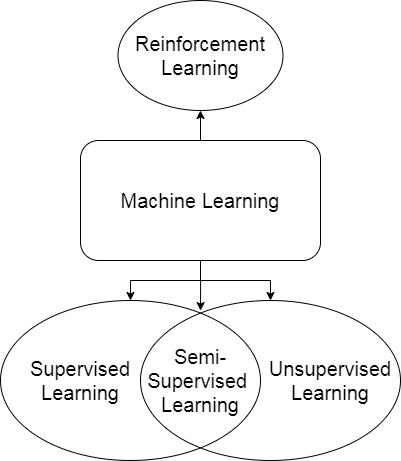}
    \caption{Machine Learning Sub-fields}
    \label{supsemi}
\end{figure}

ML algorithms are commonly classified as either deep or non-deep learning. Although DL has gained higher popularity and attention, some classical non-deep ML algorithms are more useful in certain applications, especially when data is lacking. ML algorithms can also be classified as supervised, semi-supervised, unsupervised, and RL classes, as shown in Fig.\ref{supsemi}. In this subsection, only non-RL, non-deep ML approaches are addressed; RL and DL are addressed in sections \RNum{2}.C and \RNum{2}.D, respectively.

\subsubsection{Supervised, Unsupervised and Semi-supervised Learning}

Supervised, unsupervised and semi-supervised learning are all ML approaches that can be employed to solve a broad variety of problems. 

During supervised learning, all of the training data is labeled, i.e., tagged with the correct answer. The algorithm is thus fully supervised, as it can check its predictions are right or wrong at any point in the training process. During image classification, for example, the algorithm is provided with images of different classes and each image is tagged with the corresponding class. The supervised model learns the patterns from the training data to then be able to predict labels for non-labeled data during inferencing. Supervised learning has been applied for classification and regression tasks. 

As labeling can be impossible due to a lack of information or infeasible due to high costs, unsupervised learning employs an unlabeled data set during training. Using unlabeled data, the model can extract hidden patterns or structures in the data that may be useful to understand a certain phenomenon or its output could be used as an input for other models. Unsupervised learning has been commonly used for clustering, anomaly detection, association and autoencoders (AEs). 

As a middle ground between supervised and unsupervised learning, semi-supervised learning allows a mixture of non-labelled and labaled portions of training data. Semi-supervised learning is thus an excellent option when only a small part of the data is labeled and/or the labeling process is either difficult or expensive. An example of this technique is pseudo-labeling, which has been used to improve supervised models \cite{d6}.

\subsubsection{Probabilistic Modeling}

Probabilistic modeling as mentioned by its name, involves models using statistical techniques to analyze data and was one of the earliest forms of ML \cite{b4}. A popular example is the Naive Bayes classifier, which uses Bayes' theorem while assuming that all of the input features are independent; as they generally are not, this is a naive assumption \cite{b1}. Another popular example is logistic regression; as the algorithm for this classifier is simple, it is commonly used in the data science community. 

\subsubsection{Support Vector Machine (SVM)}
Kernel methods are a popular class of algorithms \cite{b5,b1}; where the most well-known one of them is the SVM, which aims to find a decision boundary to classify data inputs. The algorithm maps the data into a high dimensional representation where the decision boundary is expressed as a hyperplane. The hyperplane is then searched by trying to maximize the distance between the hyperplane and the nearest data points from each class in a process called maximizing the margin. Although mapping the data into a high dimensional space is theoritically straightforward, it requires high computational resources. The 'kernel trick', which is based on kernel functions \cite{b6}, is thus used to compute the distance between points without explicit computation of coordinates, thereby avoiding the computation of the coordinated of a point in a high-dimensional space. SVMs have been the state-of-the-art for classification for a fairly long time and have shown many successful applications in several scientific and engineering areas \cite{c6}. However SVMs have shown limitations when applied on large datasets. Furthermore, when the SVM is applied to perceptual problems, a feature engineering step is required to enhance the performance because it is a shallow model; this requires human expertise. Although it has been surpassed by DL algorithms, it is still useful because of its simplicity and interpretability. 

\subsubsection{Decision Trees}

\begin{figure}
    \centering
    \includegraphics[scale = 0.42]{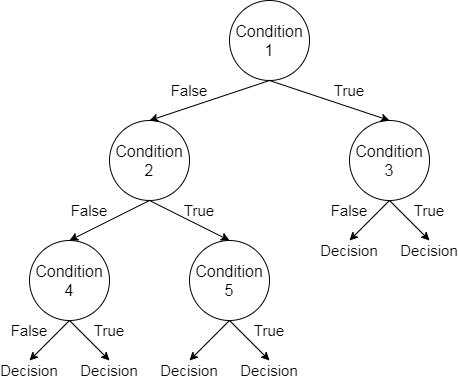}
    \caption{Decision Tree}
    \label{binary}
\end{figure} 

A decision tree is a supervised learning algorithm that represents features of the data as a tree by defining conditional control statements, as summarized in Fig.\ref{binary} \cite{a7,b7}. Given its intelligibility and simplicity, it is one of the most popular algorithms in ML. Further, decision trees can be used for both regression and classification, as decisions could be either continuous values or categories. A more robust version of decision trees, random forests (RFs), combines various decision trees to bring optimized results. This involves building many different weak decision trees and then assembling their outputs using bootstrap aggregating (bagging) \cite{c7,d7}. Another popular version of decision trees, that is often more effective than RFs is a gradient boosting machine; gradient boosting also combines various decision tree models but differs from RFs by using gradient boosting \cite{l7}, which is a way to improve ML models by iteratively training new models that focus on the mistakes of the previous models. The XGBoost \cite{e7,k7} library is an excellent implementation of the gradient boosting algorithm that supports C++, Java, Python, R, Julia, Perl, and Scala. RFs and gradient boosting machines are the most popular and robust non-deep algorithms that have been widely used to win various data science competitions on the Kaggle website \cite{f7}. 

\subsubsection{Neural Networks (NNs)}
\begin{figure}
    \centering
    \includegraphics[scale = 0.35]{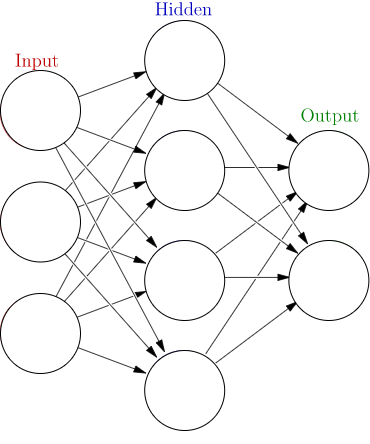}
    \caption{Neural Networks}
    \label{nn}
\end{figure} 

NNs contain different layers of interconnected nodes, as shown in Fig.\ref{nn}, where each node is a perceptron that feeds the signal produced by a multiple linear regression to an activation function that may be nonlinear \cite{a8,b8}. A nonlinear activation function is generally chosen to add more complexity to the model by eliminating linearity. NNs can be used for regression by predicting continuous values or for classification by predicting probabilities for each class. In a NN, the features of one input (e.g., one image) are assigned as the input layer. Then, according to a matrix of weights the next hidden layers are computed using matrix multiplications (linear manipulations) and then non linear activation functions. The training of NNs is all about finding the best weights. To do so, a loss function is designed to compare the output of the model and the ground truth for each output, to find the weights that minimize that loss function. Backpropagation algorithms have been designed to train chains of weights using optimization techniques such as gradient-descent \cite{c8}. NNs have been successfully used for both regression and classification, although they are most efficient when dealing a high number of features (input parameters) and hidden layers, which has led to the development of DL. 

\subsection{Deep Learning (DL)}
In contrast to shallow models, this sub-field of ML requires high computational resources \cite{d8,b1}. Recent computational advancements and the automation of feature engineering have paved the way for DL algorithms to surpass classical ML algorithms for solving complex tasks, especially perceptual ones such as computer vision and natural language processing. Due to their relative simplicity, shallow ML algorithms, require human expertise and intervention to extract valuable features or to transform the data to make it easier for the model to learn. DL models minimize or eliminate these steps as these transformations are implicitly done within the deep networks. 

\subsubsection{Convolutional Neural Networks (CNN)}


CNN \cite{d9,d10}, are a common type of deep NNs (DNNs) that are composed of an input layer, hidden convolution layers, and an output layer and have been commonly used in computer vision applications such as image classification \cite{d11}, object detection \cite{d12}, and object tracking \cite{d13}. 
They have also  shown success in other fields including speech and natural language processing \cite{d14}. As their name indicates, CNNs are based on convolutions. The hidden layers of a CNN consist of a series of convolutional layers that convolve. An activation function is chosen and followed by additional convolutions. CNN architectures are defined by by choosing the sizes, numbers, and positions of filters (kernels) and the activation functions. Learning then involves finding the best set of filters that can be applied to the input to extract useful information and predict the correct output. 

\subsubsection{Recurrent Neural Networks (RNNs)}

\begin{figure}
    \centering
    \includegraphics[scale = 0.3]{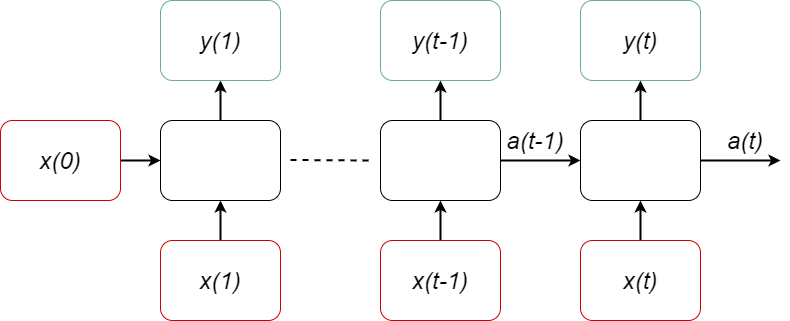}
    \caption{Simplified Architecture of a Recurrent Neural Networks}
    \label{rnn}
\end{figure} 

RNNs \cite{d15} are another family of neural networks in which nodes form a directed graph along a temporal sequence where previous outputs are used as inputs. RNNs are specialized for processing a sequence of values x(0), x(1), x(2), ..., x(T). RNNs use their internal memory to process variable-length sequences of inputs. Different architectures are designed based on the problem and the data. In general, RNNs are designed as in Fig. \ref{rnn}, where for each time stamp $t$, $x(t)$ represents the input at that time, $a(t)$ is the activation, and $y(t)$ is the output, $W_{a}$, $W_{x}$, $W_{y}$, $b_{x}$ and $b_{y}$ are coefficients that are shared temporarily and $g_{1}$ and $g_{2}$ are activation functions. 

\begin{equation}
a(t) = g_{1}(W_{a}.a(t-1)+W_{x}.x(t)+b_{a})
\label{eq}
\end{equation}

\begin{equation}
y(t) = g_{2}(W_{y}.a(t)+b_{y})
\label{eq2}
\end{equation}

RNN models are most commonly used in the fields of natural language processing, speech recognition and music composition.

\subsubsection{Autoencoders (AEs)}

\begin{figure}
    \centering
    \includegraphics[scale = 0.13]{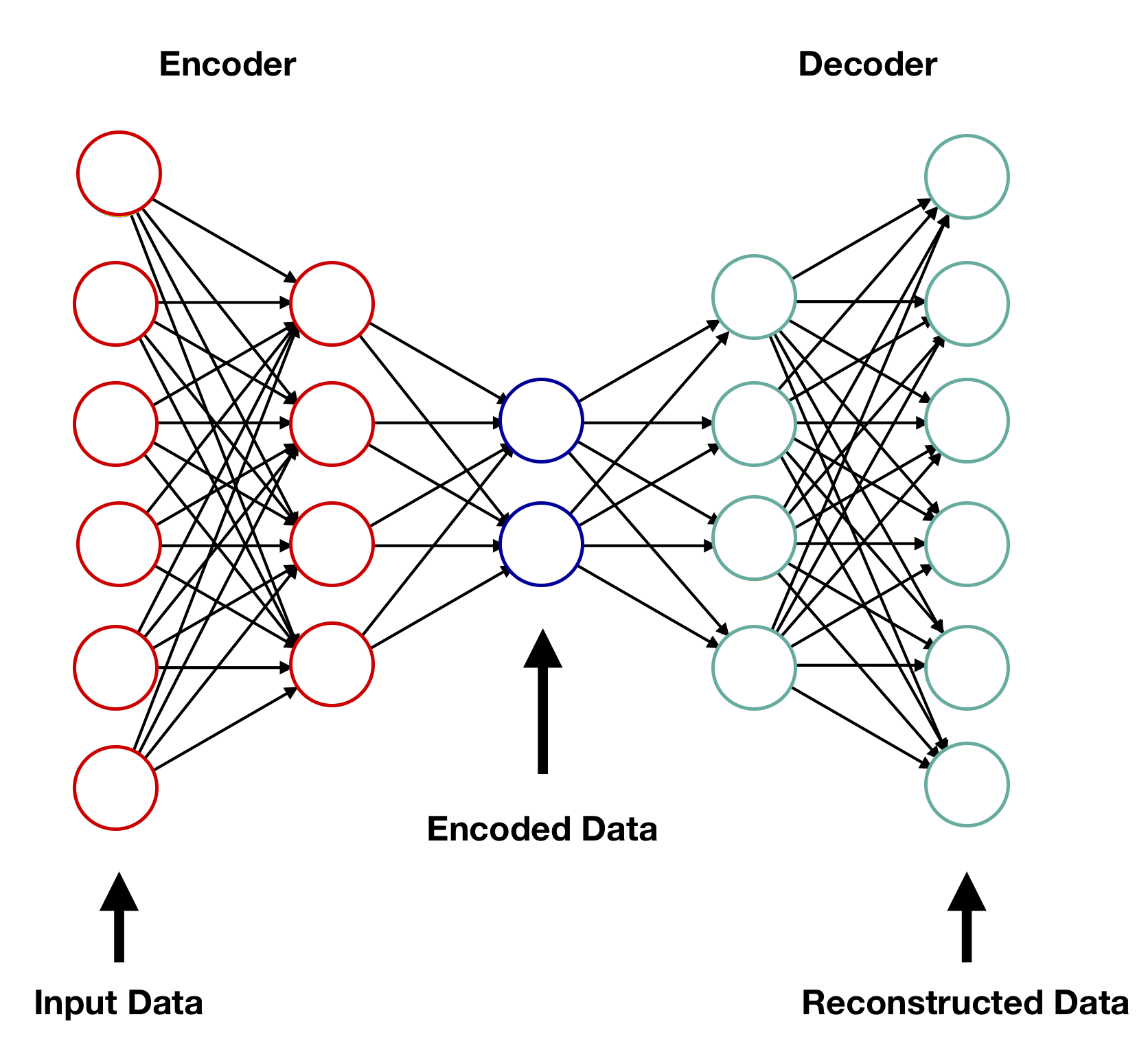}
    \caption{Autoencoder}
    \label{ae}
\end{figure} 

AEs are another type of NNs used to learn efficient data representation in an unsupervised way \cite{d16}. AEs encode the data using the bottleneck technique, which comprises dimensionality reduction to ignore the noise of the input data and an initial data regeneration from the encoded data, as summarized in Fig.\ref{ae}. The initial input and generated output are then compared to asses the quality of coding. AEs have been widely applied for for dimensionality reduction \cite{d17} and anomaly detection \cite{d18}.

\subsubsection{Deep generative models}
Deep generative models \cite{d19} are DL models that involve the automatic discovering and learning of regularities in the input data in such a way that new samples can be generated. These models have shown various applications, especially in the field of computer vision. The most popular generative models are variational AEs (VAEs) and generative adversarial networks (GANs). 

Of these, VAEs learn complicated data distribution using unsupervised NNs \cite{d20}. Although VAEs are a type of AEs, their encoding distribution is regularized during the training to ensure that their latent space (i.e., representation of compressed data) has good properties for generating new data. 

\begin{figure}
    \centering
    \includegraphics[scale = 0.25]{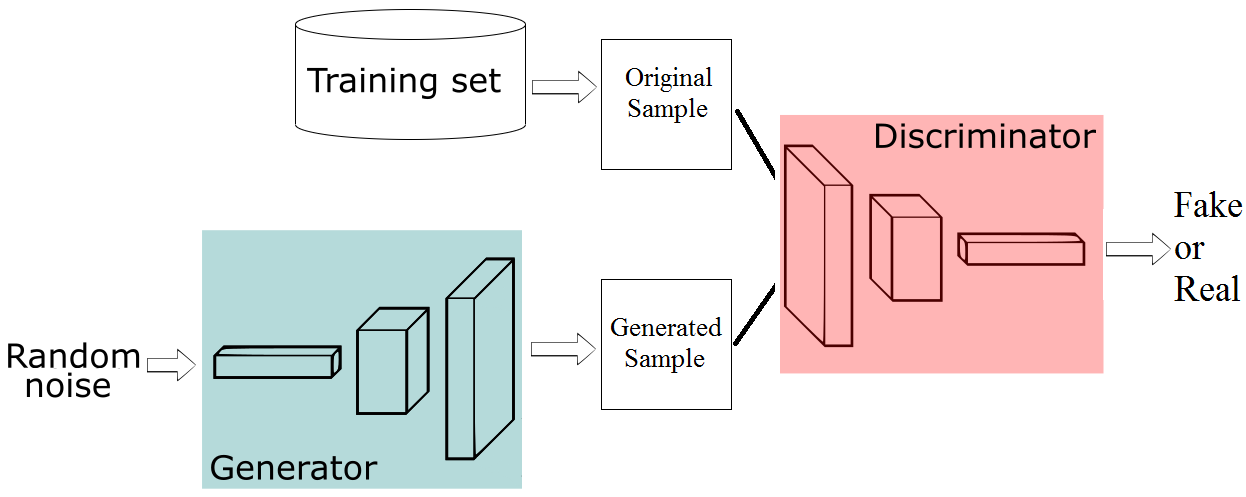}
    \caption{Generative Adverserial Networks GANs}
    \label{gans}
\end{figure} 

GANs are composed of two NNs in competition, where a generator network G learns to capture the data distribution and generate new data and a discriminator model D estimates the probability that a given sample came from the generator rather than the initial training data, as summarized in Fig. \ref{gans} \cite{d201,d21}. The generator thus is used to produce misleading samples and to that the discriminator can determine whether a given sample is fake or real. The generator fools the discriminator by generating almost real samples and the discriminator fools the generator by improving its discriminative capability. 

\subsection{Reinforcement Learning (RL)}
\begin{figure}
    \centering
    \includegraphics[scale = 0.35]{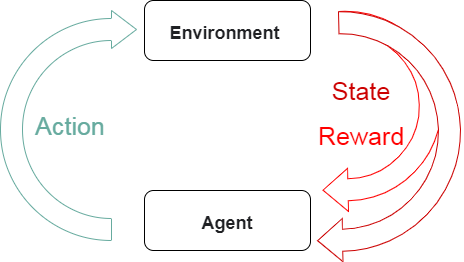}
    \caption{Reinforcement Learning}
    \label{rl}
\end{figure}

This subset of ML involves a different learning method than those using supervised, semi-supervised, or unsupervised learning \cite{b9}. RL is about learning what actions to take in the hope to maximize a reward signal. The agent must find which actions bring the most recompense by trying each action, as shown in \ref{rl}. These actions can affect immediate rewards as well as subsequent rewards. Some RL approaches require the introduction of DL; such approaches are part of deep RL (DRL).

One of the challenges encountred during RL is balancing the trade-off between exploration and exploitation. To get a maximum immediate reward, an RL agent must perform exploitation, i.e., choose actions that it has explored previously and found to be the best. To find such actions, it must explore the solution space, i.e., try new actions.  

All RL agents have explicit goals, are aware of some aspects of their environment, can take actions that impact their environments, and act despite significant uncertainty about their environment. Other than the agent and the environment, an RL system has four sub-elements: a policy, a reward signal, a value function, and, sometimes, a model of the environment.

Here, learning involves the agent determining the best method to map states of the environment to actions to be taken when in those states. After each action, the environment sends the RL agent a reward signal, which is the goal of the RL problem. Unlike a reward that brings immediate evaluation of the action, a value function estimates the total amount of recompense an agent can anticipate to collect in the longer-term. Finally, a model of the environment mimics the behavior of the environment. These models can be used for planning by allowing the agent to consider possible future situations before they occur. Methods for solving RL problems that utilize models are called model-based methods, whereas those without models are referred to as model-free methods.

\subsection{Discussion}

\subsubsection{Model Selection}

AI is a broad field that encompasses various approaches, each of which encompasses several algorithms. AI could be based on predefined rules or on ML. This learning can be supervised, semi-supervised, unsupervised, or reinforcement learning; in each of these categories learning can be deep or shallow. As each approach offers something different to the world of AI, interest in each should depend on the given problem; a more-complex approach or algorithm does not necessarily lead to better results. For example, a common assumption is that DL is better than shallow learning. Although this holds in several cases, especially for perceptual problems such as computer vision problems, it is not always applicable, as DL algorithms require greater computational resources and large datasets which are not always available. Supervised learning is an effective approach when a fully labeled dataset is available. However, this is not always the case, as data can be expensive, difficult or even impossible. Under these circumstances, semi-supervised or unsupervised learning or RL is more applicable. Whereas unsupervised learning can find hidden patterns in non-labeled data, RL learns the best policy to achieve a certain task. Thus, unsupervised learning is a good tool to extract information from data, Whereas RL is better suited for decision-making tasks. Therefore, the choice of an approach or an algorithm should not be based on its perceived elegance, but by matching the method to characteristics of the problem at hand, including the goal, the quality of the data, the computational resources, the time constraints, and the prospective future updates. Solving a problem may require a combination of more than one approach.

After assessing the problem and choosing an approach, an algorithm must be chosen. Although ML has mathematical foundations, it remains an empirical research field. To choose the best algorithm, data science and ML researchers and engineers empirically compare different algorithms for a given problem. Algorithms are compared by splitting the data into a training set and a test set. The training set is then used to train the model, whereas the test set is to compare the output between models.

In competitive data science, such as in Kaggle \cite{f7} competitions, where each incrementation matters, models are often combined to improve their overall results, and various ensemble techniques such as bagging \cite{d7}, boosting \cite{l7}, and adaptive boosting \cite{d22} are used.

\subsubsection{Model Regularization}

\begin{figure}
    \centering
    \includegraphics[scale = 0.48]{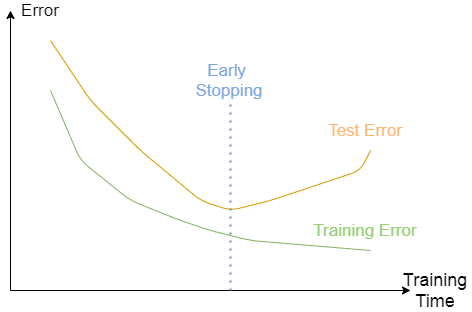}
    \caption{Training and test errors over the training time. Early stopping is common technique to reduce overfitting by stopping the training process at an early stage, i.e. when the test error starts to remarkably increasing}
    \label{over}
\end{figure} 

After the approach and algorithm have been selected, hyperparameter tuning is generally done to improve the output of the algorithm. In most cases, ML algorithms depend on many hyperparameters; choosing the best hyperparameters for a given problem thus allows for higher accuracy. This step can be done manually by intuitively choosing better hyperparameters, or automatically using various methods such as grid search and stochastic methods \cite{d23}.

A common trap in ML is overfitting, during which the machine stops learning (generalizing) and instead begins to memorize the data. When this occurs, the model can achieve good results on seen data but fails when confronted with new data, i.e., a decreased training error and an increasing test error, as shown in Fig. Fig.\ref{over}. Overfitting can be discovered by splitting the data into training, validation and testing sets, where neither the validation nor the testing sets are used to train the model. The training set is used to train the model, the validation set is used to verify the model predictions on unseen data and for hyperparameter tuning, and the testing set is used for the final testing of the model. 

A variety of methods can be employed to reduce overfitting. It be reduced by augmenting the size of the dataset, which is commonly performed in the field of computer vision. For example, image data could be augmented by applying transformations to the images, such as rotating, flipping, adding noise, or cutting parts of the images. Although useful, this technique is not always applicable. Another method involves using cross-validation  rather than splitting the data into a training set and a validation set Early stopping, as shown in Fig.\ref{over}, consists of stopping the learning process before the algorithm begins to memorize the data. Ensemble learning is also commonly used.

\subsubsection{The hype and the hope}
Rapid progress has been made in AI research, including its various subfields, over the last ten years as a result of exponentially increasing investments. However, few substantial developments have been made to address real-world problems; as such, many are doubtful that AI will have much influence on the state of technology and the world. Chollet \cite{b1} compared the progress of AI with that of the internet in 1995, the majority of people could not foresee the true potential, consequences, and pertinence of the internet, as it had yet to come to pass. As the case with the overhyping and subsequent funding crash throughout the early 2000s before the widespread implementation and application of the internet, AI may also become an integral part of global technologies. The authors thus believe that the inevitable progress of AI is likely to have long-term impacts and that AI will likely be a major part of diverse applications across all scientific fields, from mathematics to satellite communication.

\section{Artificial Intelligence for Satellite Communication}

\subsection{Beam hopping}

\begin{figure}
    \centering
    \includegraphics[scale = 0.5]{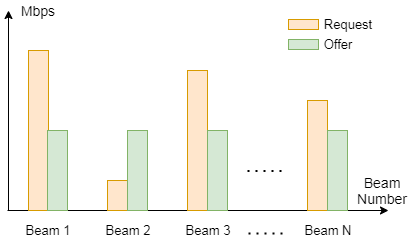}
    \caption{The demand–capacity mismatch among beams demonstrates the limitation of using fixed and uniformly distributed resources across
all beams in a multi-beam satellite system}
    \label{bh}
\end{figure} 

\begin{figure}
    \centering
    \includegraphics[scale = 0.42]{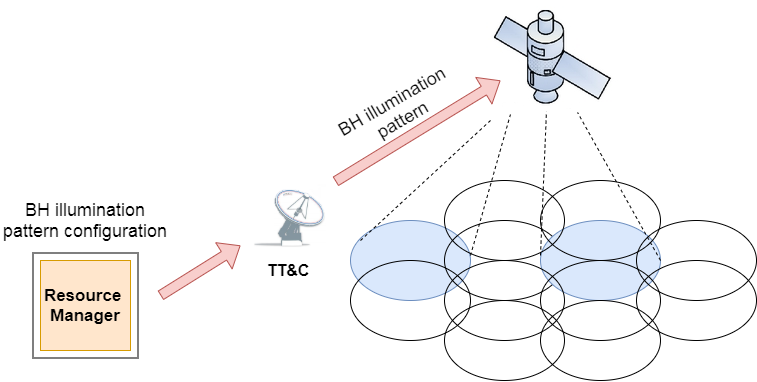}
    \caption{Simplified architecture of beam hopping (BH)}
    \label{multi}
\end{figure} 

\subsubsection{Definition \& limitations}
Satellite resources are expensive and thus require efficient systems involving optimizing and time-sharing. In conventional satellite systems the resources are fixed and uniformly distributed across beams \cite{bh1}. As a result, conventional large multi-beam satellite systems have shown a mismatch between the offered and requested resources; some spot beams have a higher demand than the offered capacity, leaving the demand pending (i.e., hot-spots), while others present a demand lower than the installed capacity, leaving the offered capacity unused (i.e., cold-spots). Thus, to improve multi-beam satellite communication, the on-board flexible allocation of satellite resources over the service coverage area is necessary to achieve more efficient satellite communication. 

Beam hopping (BH) has emerged as a promising technique to achieve greater flexibility in managing non-uniform and variant traffic requests throughout the day, year and lifetime of the satellite over the coverage area \cite{bh1}, \cite{bh2}. BH, involves dynamically illuminating each cells with a small number of active beams, as summarized in \ref{multi}, thus using all available on-board satellite resources to offer service to only a subset of beams. The selection of this subset is time-variant and depends on the traffic demand, which is based on the time-space dependent BH illumination pattern. The illuminated beams are only active long enough to fill the request for each beam. Thus, the challenging task in BH systems is to decide which beams should be activated and for how long, i.e., the BH illumination pattern; this responsibility is left to the resource manager who then forwards the selected pattern to the satellite via telemetry, tracking and command \cite{bh3}.

Of the various methods that researchers have provided to realize BH, most have been based on classical optimization algorithms. For example, Angeletti et al. \cite{bhst1}, demonstrated several advantages to the performance of a system when using BH and proposed the use of genetic algorithm (GA) to design the BH illumination pattern; Anzalchi et al. \cite{bhst2}, also illustrated the merits of BH and compared the performance between BH and non-hopped systems. Alberti et al. \cite{bhst25}, proposed a heuristic iterative algorithm to obtain a solution to the BH illumination design. BH has also been used to decrease the number of transponder amplifiers for Terabit/s satellites \cite{bhst3}. An iterative algorithm has also been proposed to maximize the overall offered capacity under certain beam demand and power constraints in a joint BH design and spectrum assignment \cite{bhst4}. Alegre et al. \cite{bhst5}, designed two heuristics to allocate capacity resources basing on the traffic request per-beam, and then further discussed the long and short-term traffic variations and suggested techniques to deal with both variations \cite{bhst6}. Liu et al. \cite{bhst7}, studied techniques for controlling the rate of the arriving traffic in BH systems. The QoS delay fairness equilibrium has also been addressed in BH satellites \cite{bhst8}. Joint BH schemes were proposed  by Shi et al. \cite{bhst9} and Ginesi et al. \cite{bhst10} to further ameliorate the efficiency of on-board resource allocation. To find the optimal BH illumination design, Cocco et al. \cite{bhst11} used a simulated annealing algorithm.

Although employing optimization algorithms has achieved satisfactory results in terms of flexibility and delay reduction of BH systems, some difficulties remain. As the search space dramatically grow with the number of beams, an inherent difficulty in designing the BH illumination pattern is finding the optimal design rather than one of many local optima \cite{bhst4}. For satellites with hundreds or thousands of beams, classical optimization algorithms may require long computation times which is impractical in many scenarios. 

Additionally, classical optimization algorithms, including the GAs or other heuristics, require revision when the scenario changes moderately; this leads to a higher computational complexity, which is impractical for on-board resource management. 

\subsubsection{AI-based solutions}
Seeking to overcome these limitations and enhance the performance of BH, some researchers have proposed AI-based solutions. Some of these solutions have been fully based on the learning approach, i.e., end-to-end learning, in which the BH algorithm is a learning algorithm. Others have tried to improve optimization algorithms by adding a learning layer, thus combining learning and optimization. 

To optimize the transmission delay and the system throughput in multibeam satellite systems, Hu et al \cite{bhst12} formulated an optimization problem and modeled it as a Markov decision process (MDP). DRL is then used to solve the BH illumination design and optimize the long-term accumulated rewards of the modeled MDP. As a result, the proposed DRL-based BH algorithm can reduce the transmission delay by up to 52.2\% and increased the system throughput by up to 11.4\% when compared with previous algorithms.

To combine the advantages of end-to-end learning approaches and optimization approaches, for a more efficient BH illumination pattern design, Lei et al. \cite{bh3} suggested a learning and optimization algorithm to deal with the beam hopping pattern illumination selection, in which a learning approach, based on fully connected NNs, was used to predict non-optimal BH patterns and thus address the difficulties faced when applying an optimization algorithm to a large search space. Thus, the learning-based prediction reduces the search space, and the optimization can be reduced on a smaller set of promising BH patterns.

Researchers have also employed multi-objective DRL (MO-DRL) for the DVB-S2X satellite. Under real conditions, Zhang et al. \cite{bhst13} demonstrated that the low-complexity MO-DRL algorithm could ensure the fairness of each cell, and ameliorate the throughput better than previous techniques including DRL \cite{bhst11} by 0.172\%. In contrast, the complexity of GA producing a similar result is about 110 times that of the MO-DRL model. Hu et al. \cite{bhst14} proposed a multi-action selection technique based on double-loop learning and obtained a multi-dimensional state using a DNN. Their results showed that the proposed technique can achieve different objectives simultaneously, and can allocate resources intelligently by adapting to user requirements and channel conditions.

\subsection{Anti-jamming}
\subsubsection{Definition \& limitations}
Satellite communication systems are required to cover a wide area, and provide high-speed, communication and high-capacity transmission. However, in tactical communication systems using satellites, reliability and security are the prime concerns; therefore, an anti-jamming (AJ) capability is essential. Jamming attacks could be launched toward main locations and crucial devices in a satellite network to reduce or even paralyze the throughput. Several AJ methods have thus been designed to reduce possible attacks and guarantee secure  satellite communication.

The frequency-hopping (FH) spread spectrum method has been preferred in many prior tactical communication systems using satellites \cite{aj1,aj2}. Using the dehop–rehop transponder method employing FH-frequency division multiple access (FH-FDMA) scenarios, Bae et al. \cite{aj3} developed an efficient synchronization method with an AJ capability.

Most prior AJ techniques are not based on learning and thus cannot deal with clever jamming techniques that are capable of continuously adjusting the jamming methodology by interaction and learning. Developing AI algorithms offer advanced tools to achieve diverse and intelligent jamming attacks based on learning approaches and thus present a serious threat to satellite communication reliability. In two such examples, a smart jamming formulation automatically adjusted the jamming channel \cite{aj4}, whereas a smart jammer maximized the jamming effect by adjusting both the jamming power and channel \cite{aj5}. In addition, attacks could be caused by multiple jammers simultaneously implementing intelligent jamming attacks based on learning approaches. Although this may be an unlikely scenario, it has not yet been seriously considered. Further, most researchers have focused on defending against AJ attacks in the frequency-based domain, rather than spacebased AJ techniques, such as routing AJ.

\subsubsection{AI-based solutions}
\begin{figure}
    \centering
    \includegraphics[scale = 0.35]{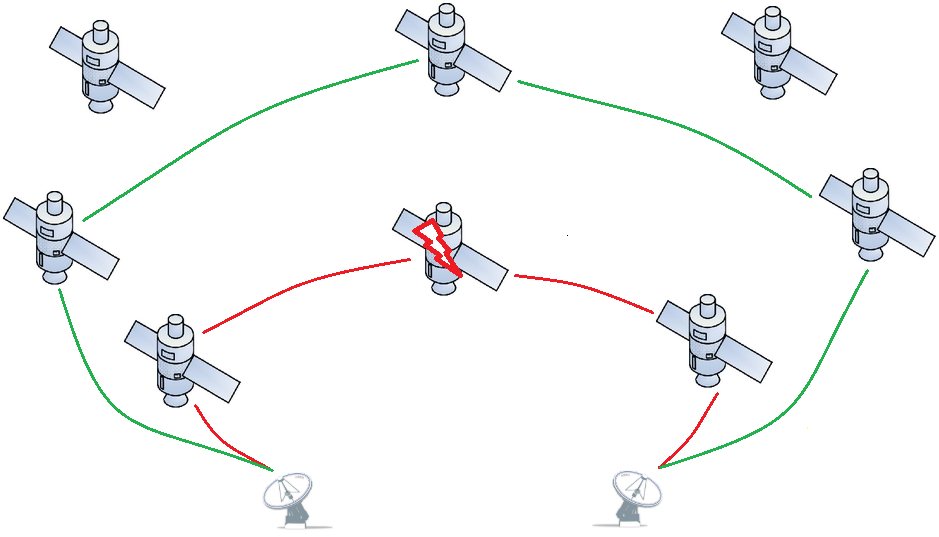}
    \caption{Space-based anti-jamming (AJ) routing. The red line represents the found jammed path, and the green one represents the suggested path \cite{aj7}}
    \label{aj}
\end{figure}

By using a long short-term memory (LSTM) network, which is a DL RNN, to learn the temporal trend of a signal, Lee et al. \cite{aj6} demonstrated a reduction of overall synchronization time in the previously discussed FH-FDMA scenario \cite{aj3}. Han et al. \cite{aj7} proposed the use of a learning approach for AJ to block smart jamming in the Internet of Satellites (IoS) using a space-based AJ method, AJ routing, summarized in Fig.\ref{aj}. By combining game theory modeling with RL and modeling the interactions between smart jammers and satellite users as a Stackelberg AJ routing game, they demonstrated how to use DL to deal with the large decision space caused by the high dynamics of the IoS and RL to deal with the interplay between the satellites and the smart jamming environment. DRL thus made it possible to solve the routing selection issue for the heterogeneous IoS while preserving an available routing subset to simplify the decision space for the Stackelberg AJ routing game. Based on this routing subset, a popular RL algorithm, Q-Learning, was then used to respond rapidly to intelligent jamming and adapt AJ strategies.

Han et al. \cite{aj8} later combined game theory modeling and RL to obtain AJ policies according to the dynamic and unknown jamming environment in the Satellite-Enabled Army IoT (SatIoT). Here, a distributed dynamic AJ coalition formation game was examined to decrease the energy use in the jamming environment, and a hierarchical AJ Stackelberg game was proposed to express the confrontational interaction between jammers and SatIoT devices. Finally, RL-based algorithms were utilized to get the sub-optimal AJ policies according to the jamming environment. 

\subsection{Network Traffic Forecasting}
\subsubsection{Definition \& limitations}
Network traffic forecasting is a proactive approach that aims to guarantee reliable and high-quality communication, as the predictability of traffic is important in many satellite applications, such as congestion control, dynamic routing, dynamic channel allocation, network planning, and network security. Satellite network traffic is self-similar and demonstrates long-range-dependence (LRD) \cite{tf1}. To achieve accurate forecasting, it is therefore necessary to consider its self-similarity. However,forecasting models for terrestrial networks based on self-similarity have a high computational complexity; as the on-board satellite computational resources are limited, terrestrial models are not suitable for satellites. An efficient traffic forecasting design for satellite networks is thus required.

Several researchers have performed traffic forecasting for both terrestrial and satellite networks; these techniques have included the Markov \cite{tf2}, autoregressive moving average  (ARMA) \cite{tf3}, autoregressive integrated moving average (ARIMA) \cite{tf4} and fractional ARINA (FARIMA) \cite{tf5} models. By using empirical mode decomposition (EMD) to decompose the network traffic and then applying the ARMA forecasting model, Gao et al. \cite{tf6} demonstrated remarkable improvement. 

The two major difficulties facing satellite traffic forecasting are the LRD of satellite networks and the limited on-board computational resources. Due to the LRD property of satellite networks, short-range-dependence (SRD) models have failed to achieve accurate forecasting. Although previous LRD models have achieved better results than SRD models, they  suffer from high complexity. To address these issues, researchers have turned to AI techniques.

\subsubsection{AI-based solutions}
Katris and Daskalaki \cite{tf5} combined FARIMA with NNs for internet traffic forecasting, whereas Pan et al. \cite{tf7} combined a differential evolution with NNs for network traffic prediction. Due to the high complexity of classical NNs, a least-square SVM, which is an optimized version of a SVM, has also been used for forecasting \cite{tf8}. By applying principal component analysis (PCA), to reduce the input dimensions and then a generalized regression NN, Ziluan and Xin \cite{tf9} achieved higher-accuracy forecasting with less training time. Zhenyu et al. \cite{tf10} used traffic forecasting as a part of their distributed routing strategy for LEO satellite network. An extreme learning machine (ELM) has also been employed for traffic load forecasting of satellite node before routing \cite{tf11}. Bie et al. \cite{tf1} used EMD to decompose the traffic of the satellite with LRD into a series with SRD and at one frequency to decrease the predicting complexity and augment the speed. Their combined EMD, fruit-fly optimization, and ELM methodology achieved more accurate forecasting at a higher speed than prior approaches.

\subsection{Channel Modeling}
\subsubsection{Definition \& limitations}
A channel model is a mathematical representation of the effect of a communication channel through which wireless signals are propagated; it is modeled as the impulse response of the channel in the frequency or time domain. 

A wireless channel presents a variety of challenges for reliable high-speed communication, as it is vulnerable to noise, interference, and other channel impediments, including path loss and shadowing. Of these, path loss is caused by the waste of the power emitted by the transmitter and the propagation channel effects, whereas shadowing is caused by the obstacles between the receiver and transmitter that absorb power \cite{ch1}. 
 
Precise channel models are required to asses the performance of mobile communication system and therefore to enhance coverage for existing deployments. Channel models may also be useful to forecast propagation in designed deployment outlines, which could allow for assessment before deployment, and for optimizing the coverage and capacity of actual systems. 
For small number of transmitter possible positions, outdoor extensive environment evaluation could be done to estimate the parameters of the channel \cite{ch2,ch3}. As more advanced technologies have been used in wireless communication, more advanced channel modelling was required. Therefore the use of stochastic models that are computationally efficient while providing satisfactory results \cite{ch4}.

Ray tracing is used for channel modeling, which requires 3D images that are generally generated using computer vision methods including stereo-vision-based depth estimation \cite{ch45,ch46}, \cite{ch6,ch7}.

A model is proposed for an urban environment requires features, including road widths, street orientation angles, and height of buildings \cite{ch8}. A simplified model was then proposed, by Fernandes and Soares \cite{ch9} that required only the proportion of building occupation between the receiver and transmitter, which could be computed from segmented images manually or automatically \cite{ch10}.

Despite the satisfactory performance of some of the listed techniques, they still have many limitations. 
For example, the 3D images required by ray tracing r are not generally available and their generation is not computationally efficient. Even when the images are available, ray tracing is computationally costly and data exhaustive and therefore is not appropriate for real-time coverage area optimization. Further, the detailed data required for the model presented by Cichon and Kurner \cite{ch8} is often unavailable.

\subsubsection{AI-based solutions}
Some early applications of AI for path loss forecasting have been based on classical ML algorithms such as SVM \cite{ch11,ch111}, NNs \cite{ch12,ch13,ch14,ch15,ch16,ch17} and decision trees \cite{ch18}. Interested readers are referred to a survey of ML-based path loss prediction approaches for further details \cite{ch19}.

\begin{figure}
    \centering
    \includegraphics[scale = 0.43]{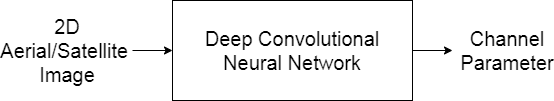}
    \caption{Channel parameters prediction. 2D aerial/satellite images used as input to the deep convolutional neural network (CNN)to to predict channel parameters. The model is trained separately for each parameter.}
    \label{ch}
\end{figure} 
 
However, although previous ML efforts have shown great results, many require 3D images. Researchers have recently thus shifted their attention to using DL algorithms with 2D satellite/aerial images for path loss forecasting. For example, Ates et al. \cite{ch20}, approximated channel parameters, including the standard deviation of shadowing and the path loss exponent, from satellite images using deep CNN without the use of any added input parameters, as shown in Fig.\ref{ch}. 

By using a DL model on satellite images and other input parameters to predict the reference signal received power (RSRP) for specific receiver locations in a specific scenario/area, Thrane et al. \cite{ch21}  demonstrated a gain improvement of $\approx 1$ and $\approx 4.7$ at 811 MHz and 2630 MHz respectively, over previous techniques, including ray tracing. Similarly Ahmadien et al. \cite{ch22}, applied DL on satellite images for path loss prediction, although they focused only on satellite images without any supplemental features and worked on more generalized data. Despite the practicality of this method, as it only needs satellite images to forecast the path loss distribution, 2D images will not always be sufficient to characterize the 3D structure. In these cases, more features (e.g., building heights) must be input into the model.

\subsection{Telemetry Mining}
\subsubsection{Definition \& limitations}
Telemetry is the process of recording and transferring measurements for control and monitoring. In satellite systems, on-board telemetry helps mission control centers track platform's status, detect abnormal events, and control various situations. 
 
Satellite failure can be caused by a variety of things; most commonly, failure is due to the harsh environment of space, i.e., heat, vacuum, and radiation. The radiation environment can affect critical components of a satellite, including the communication system and power supply.
 
Telemetry processing enables tracking of the satellite's behavior to detect and minimize failure risks. Finding correlations, recognizing patterns, detecting anomalies, classifying, forecasting, and clustering are applied to the acquired data for fault diagnosis and reliable satellite monitoring.

One of the earliest and simplest techniques used in telemetry analysis is limit checking. The method is based on setting a precise range for each feature (e.g., temperature, voltage, and current), and then monitoring the variance of each feature to detect out-of-range events. The main advantage of this algorithm is its simplicity limits, as can be chosen and updated easily to control spacecraft operation. 

Complicated spacecraft with complex and advanced applications challenges current space telemetry systems. Narrow wireless bandwidth and fixed-length frame telemetry make transmitting the rapidly augmenting telemetry volumes difficult. In addition, the discontinuous short-term contacts between spacecraft and ground stations limit the data transmission capability. Analyzing, monitoring and interpreting huge telemetry parameters could be impossible due to the high complexity of data.

\subsubsection{AI-based solutions}
In recent years, AI techniques have been largely considered in space missions with telemetry. Satellite health monitoring has been performed using probabilistic clustering \cite{tel1}, dimensionality reduction, and hidden Markov \cite{tel2}, and regression trees \cite{tel3}, whereas others have developed anomaly detection methods using the K-nearest neighbor (kNN), SVM, LSTM and testing on the telemetry of Centre National d’Etudes Spatiales spacecraft \cite{tel4,tel5,tel6}.

Further, the space functioning assistant was further developed in diverse space applications using data-driven \cite{tel7} and model-based \cite{tel8} monitoring methods. In their study of the use of AI for fault diagnosis in general and for space utilization, Sun et al. \cite{tel9} argued that the most promising direction is the use of DL; suggested its usage for fault diagnosis for space utilization in China.
 
By comparing different ML algorithms using telemetry data from the Egyptsat-1 satellite, Ibrahim et al. \cite{tel10} demonstrated the high prediction accuracy of LSTM, ARIMA, and RNN models. They suggested simple linear regression for forecasting critical satellite features for short-lifetime satellites (i.e., 3–5 years) and NNs for long-lifetime satellites (15-20 years).
 
Unlike algorithms designed to operate on the ground in the mission control center, Wan et al.
\cite{tel11} proposed a self-learning classification algorithm to achieve on-board telemetry data classification with low computational complexity and low time latency.

\subsection{Ionospheric Scintillation Detecting}
\subsubsection{Definition \& limitations}

\begin{figure}
    \centering
    \includegraphics[scale = 0.18]{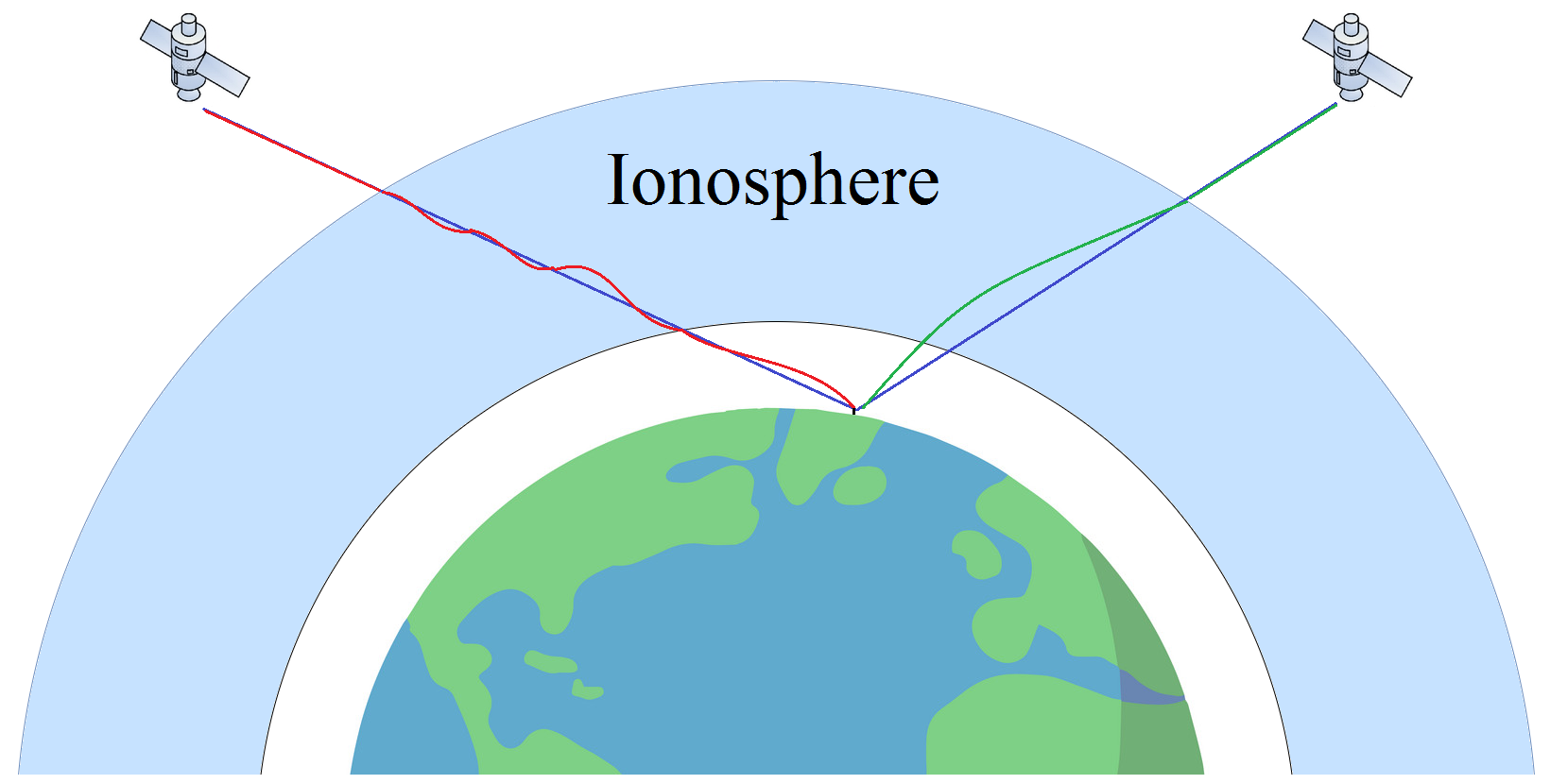}
    \caption{Representation of ionospheric scintillation, where distortion occurs during signal propagation. The blue, green, and red lines show the line-of-sight signal paths from the satellite to the earth antennas, the signal fluctuation, and the signal delay, respectively.}
    \label{ion}
\end{figure}

Signals transmission by satellites toward the earth can be notably impacted due to their propagation through the atmosphere, especially the ionosphere, which is the ionized part of the atmosphere higher layer, and is distinguished by an elevated density of free electrons (Fig.\ref{ion}). The potential irregularities and gradients of ionization can distort the signal phase and amplitude, in a process known as ionospheric scintillation.
 
In particular, propagation through the ionosphere can cause distortion of global navigation satellite system (GNSS) signals, leading to significant errors in the GNSS-based applications. GNSSs are radio-communication satellite systems that allow a user to compute the local time, velocity, and position in any place on the Earth by processing signals transferred from the satellites and conducting trilateration \cite{ion21}. GNSSs can also be used in a wide variety of applications, such as scientific observations.

Because of the low-received power of GNSS waves, any errors significantly threaten the accuracy and credibility of the positioning systems. GNSS signals propagating through the ionosphere face the possibility of both a temporal delay and scintillation. Although delay compensation methods are applied to all GNSS receivers \cite{ion21}, scintillation is still a considerable issue, as its quasi-random nature makes it difficult to model \cite{ion22}. Ionospheric scintillation thus remains a major limitation to high-accuracy applications of GNSSs. The accurate detection of scintillation thus required to improve the credibility and quality of GNSSs \cite{ion5}. To observe the signals, which are a source of knowledge for interpreting and modeling the atmosphere higher layers, and to raise caution and take countermeasures for GNSS-based applications, networks of GNSS receivers, have been installed, both at high and low latitudes, where scintillation is expected to occur \cite{ion6,ion7}. Robust receivers and proper algorithms for scintillation-detecting algorithms are thus both required \cite{ion9}.

To evaluate the magnitude of scintillation impacting a signal, many researchers have employed simple event triggers, based on the comparison of the amplitude and phase of two signals over defined interval \cite{ion12}. Other proposed alternatives, have included using wavelet techniques \cite{ion13}, decomposing the carrier-to-noise density power propostion via adaptive frequency-time techniques \cite{ion14}, and assessing the histogram statistical properties of collected samples \cite{ion15}. 

Using simple predefined thresholds to evaluate the magnitude of scintillation can be deceptive due its complexity. The loss of the transient phases of events could cause a delay in raising possible caution flags, and weak events with high variance could be missed. Further, it can be difficult to distinguish between signal distortions caused by other phenomena, including multi-path. However, other proposed alternatives depend on complex and computationally costly operations or on customized receiver architectures.

\subsubsection{AI-based solutions}
Recently, studies have proved that AI can be utilized for the detection of scintillation. For example, Rezende et al. \cite{ion16}, proposed a survey of data mining methods, that rely on observing and integrating GNSS receivers. 
 
A technique based on the SVM algorithm has been suggested for amplitude scintillation detection \cite{ion17,ion18}, and then later expanded to phase scintillation detection \cite{ion19,ion20}. 

By using decision trees and RF to systematically detect
ionospheric scintillation events impacting the amplitude of the GNSS signals, Linty et al.’s \cite{ionnew1} methodology outperformed state-of-the art methodologies in terms of accuracy (99.7\%) and F-score (99.4\%), thus reaching the levels of a manual human-driven annotation.
 
More recently, Imam and Dovis \cite{ionnew2} proposed the use of decision trees, to differentiate between ionospheric scintillation and multi-path in GNSS scintillation data. Their model, which annotates the data as scintillated, multi-path affected, or clean GNSS signal, demonstrated an accuracy of 96\%

\subsection{Managing Interference}
\subsubsection{Definition \& limitations}
Interference managing is mandatory for satellite communication operators, as interference negatively affects the communication channel, resulting in a reduced QoS, lower operational efficiency and loss of revenue \cite{in1}. Moreover, interference is a common event that is increasing with the increasing congestion of the satellite frequency band as more countries are launching satellites and more applications are expected. With the growing number of users sharing the same frequency band, the possibility of interfering augments, as does the risk of intentional interference, as discussed in section III.B. 

Interference managing is a thus essential to preserve high-quality and reliable communication systems; management includes detection, classification, and suppression of interference, as well as the application of techniques to minimize its occurrence.

Interference detection is a well-studied subject that has been addressed in the past few decades \cite{in2,in3}, especially for satellite communication \cite{in1,in31}.

However, researchers have commonly relied on the decision theory of hypothesis testing, in which specific knowledge of the signal characteristics and the channel model is needed. Due, to the contemporary diverse wireless standards, the design of specific detectors for each signal category is fruitless approach.

\subsubsection{AI-based solutions}

\begin{figure}
    \centering
    \includegraphics[scale = 0.68]{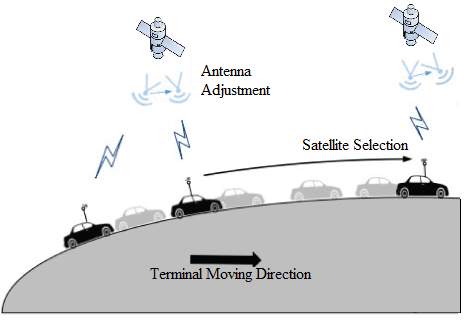}
    \caption{Satellite selection and antenna adjustment}
    \label{tr}
\end{figure} 

To minimize interference, Liu et al. \cite{in6}, suggested the use of AI for moving terminals and stations in satellite-terrestrial networks by proposing a framework combining different AI approaches including SVM, unsupervised learning and DRL for satellite selection, antenna pointing and tracking, as summarized in Fig.\ref{tr}.

Another AI-based approach executes automatic real-time interference detection is based on the forecasting of the following signal spectrum to be received in absence of anomaly, by using LSTM trained on historical anomaly-free spectra \cite{in4}. Here the predicted spectra is then compared to the received signal using a designed metric, to detect anomalies.

Henarejos et al. \cite{in5} proposed the use of two AI-based approaches, DNN AEs and LSTM, for detecting and classifying interference, respectively. In the former, the AE is trained with interference free signals and tested against other signals without interference to obtain practical thresholds. The difference in error in signals with and without interference is then exploited to detect the presence of interference.

\subsection{Remote sensing (RS)}
\subsubsection{Definition \& limitations}

RS is the process of extracting information about an area, object or phenomenon by processing its reflected and emitted radiation at a distance, generally from satellite or aircraft. 
 
RS has a wide range of applications in multiple fields including land surveying, geography, geology, ecology, meteorology, oceanography, military and communication. As RS offers the possibility of monitoring areas that are dangerous, difficult or impossible to access, including mountains, forests, oceans and glaciers it is a popular and active research area.

\subsubsection{AI-based solutions}
The revolution in computer vision capabilities caused by DL has led to the
increased development of RS by adopting state-of-the-art DL algorithms on satellite images, image classification for RS has become most popular task in computer vision. For example, Kussul et al. \cite{rs1} used DL to classify land coverage and crop types using RS images from Landsat-8 and Sentinel-1A over a test site in Ukraine. Zhang et al \cite{rs2} combined DNNs by using a gradient-boosting random CNN for scene classification. More recently, Chirayath et al. \cite{rss3} proposed the combination of kNN and CNN to map coral reef marine habitats worldwide with RS imaging. RS and AI have also been used in communication theory applications, such as those discussed in section III.D \cite{ch20}, \cite{ch21} and \cite{ch22}.

Many object detection and recognition applications have been developed using AI on RS images \cite{rss4}. Recently, Zhou et al. \cite{rs3} proposed the use of YOLOv3 \cite{rs4,rs5}, a CNN-based object detection algorithm, for vehicle detection in RS images. Others have proposed the use of DL for other object detection tasks, such as, building \cite{rs6}, airplane \cite{rs7}, cloud \cite{rs8}, \cite{rs9,rs10}, ship \cite{rs11,rs12}, and military target \cite{rs13} detection. AI has also been applied to segment and restore RS images, e.g., in cloud restorations, during which ground regions shadowed by clouds are restored.

Recently, Zheng et al. \cite{rss1} proposed a two-stage cloud removal method in which U-Net \cite{rss2} and GANs are used to perform cloud segmentation and image restoration, respectively.

AI proposed for on-board scheduling of agile Earth-observing satellites,  as autonomy improves their performance and allows them to acquire more images, by relying on on-board scheduling for quick decision-making. By comparing the use of RF, NNs, and SVM to prior learning and non-learning-based approaches, Lu et al. \cite{rs14} demonstrated that RF improved both the solution quality and response time.

\subsection{Behavior Modeling}
\subsubsection{Definition \& limitations}
Owing to the increasing numbers of active and inactive (debris) satellites of diverse orbits, shapes, sizes, orientations and functions, it is becoming infeasible for analysts to simultaneously monitor all satellites. Therefore, AI, especially ML, could play a major role by helping to
automate this process.

\subsubsection{AI-based solutions}
Mital et al. \cite{bm1} discussed the potential of ML algorithms to model satellite behavior. Supervised models have been used to determine satellite stability \cite{bm11}, whereas unsupervised models have been used to detect anomalous behavior and a satellites’ location \cite{bm111}, and an RNN has been used to predict satellite maneuvers over time\cite{bm12}.

Accurate satellite pose estimation, i.e., identifying a satellite’s relative position and attitude, is critical in several space operations, such as debris removal, inter-spacecraft communication, and docking. The recent proposal for satellite pose estimation from a single image via combined ML and geometric optimization by Chen et al. \cite{bm2} won the first place in the recent Kelvins pose estimation challenge organized by the European Space Agency \cite{bm3}.

The amount of space debris has augmented immensely over the last few years, which can cause a crucial menace to space missions due to the high velocity of the debris. It is thus essential to classify space objects and apply collision avoidance techniques  to protect active satellites. As such, Jahirabadkar et al. \cite{bm4} presented a survey of diverse AI methodologies, for classification of space objects using the curves of light as a differentiating property. 

Yadava et al. \cite{bm5} employed NNs and RL for on-board attitude determination and control; their method effectively provided the needed torque to stabilize a nanosatellite along three axes.

To avoid catastrophic events because of battery failure, Ahmed et al. \cite{bm6} developed an on-board remaining battery life estimation system using ML and a logical analysis of data approaches.

\subsection{Space-Air-Ground Integrating}
\subsubsection{Definition \& limitations}

Recently, notable advances have been made in ground communication systems to provide users higher-quality internet access. Nevertheless, due to the restricted capacity and coverage area of networks, such services are not possible everywhere at all times, especially for users in rural or disaster areas.

\begin{figure}
    \centering
    \includegraphics[scale = 0.55]{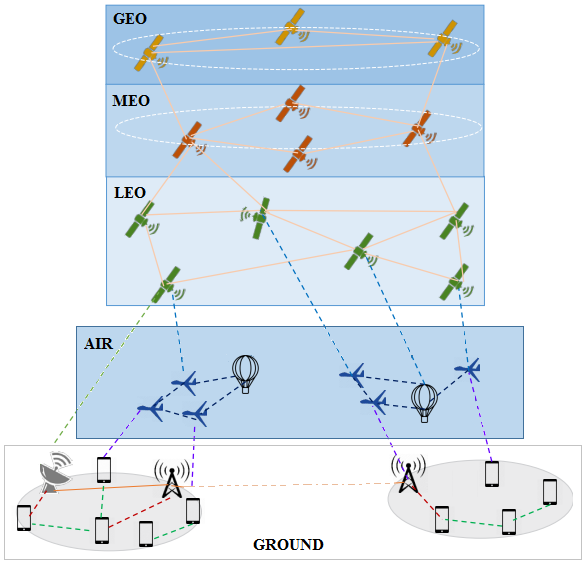}
    \caption{Space-air-ground integrated networks (SAGINs) \cite{sagin2}}
    \label{SAGIN}
\end{figure} 

Although terrestrial networks have the most resources and highest throughput, non-terrestrial communication systems have a much broader coverage area. However, non-terrestrial networks have their own limitations; e.g., satellite communication systems have a long propagation latency, and air networks have a narrow capacity and unstable links. 

To supply users with better and more-flexible end-to-end services by taking advantage of the way the networks can complement each other, researchers have suggested the use of space-air-ground integrated networks (SAGINs) \cite{sagin1}, which include the satellites in space, the balloons, airships, and UAVs in the air, and the ground segment, as shown in Fig.\ref{SAGIN}. 

The multi-layered satellite communication system which consists of GEO, MEO, and LEO satellites, can use multi-cast and broadcast methods to ameliorate the network capacity, crucially easing the augmenting traffic burden \cite{sagin1,sagin2}. As SAGINs allow packet transmission to destinations via multiple paths of diverse qualities, they can offer different packet transmissions methods to encounter diverse service demands \cite{sagin2}.

However, the design and optimization of SAGINs is more challenging than that of conventional ground communication systems owing to their inherent self-organization, time-variability, and heterogeneity \cite{sagin1}. A variety of factors that must be considered when designing optimization techniques have thus been identified \cite{sagin1,sagin2}. For example, the diverse propagation mediums, the sharing of frequency bands by different communication types, the high mobility of the space and air segments, and the inherent heterogeneity between the three segments, make the network control and spectrum management of SAGIN arduous. The high mobility results in frequent handoffs, which makes safe routing more difficult to realize, thus making SAGINs more exposed to jamming. Further, as optimizing the energy efficiency is also more challenging than in standard terrestrial networks, energy management algorithms are also required. 

\subsubsection{AI-based solutions}
In their discussion of challenges facing SAGINs, Kato et al. \cite{sagin2} proposed the use of a CNN for the routing problem to optimize the SAGIN's overall performance using traffic patterns and the remaining buffer size of GEO and MEO satellites.

Optimizing the satellite selection and the UAV location to optimize the end-to-end data rate of the Source-Satellite-UAV-Destination communication is challenging due to the vast orbiting satellites number and the following time-varying network architecture. To address this problem, Lee et al. \cite{sagin3} jointly optimized the source-satellite-UAV association and the location of the UAV via DRL. Their suggested technique achieved up to a 5.74x higher average data rate than a direct communication baseline in the absence of UAV and satellite.
 
For offloading calculation-intensive applications, a SAGIN edge/cloud computing design has been developed in such a way that satellites give access to the cloud and UAVs allow near-user edge computing. \cite{sagin4}. Here, a joint resource allocation and task scheduling approach is used to allocate the computing resources to virtual machines and schedule the offloaded tasks for UAV edge servers, whereas an RL-based computing offloading approach handles the multidimensional SAGIN resources and learns the dynamic network conditions. Here, a joint resource allocation and task scheduling approach is used to assign the computing resources to virtual machines and plan the offloaded functions for UAV edge servers, whereas an RL-based computing offloading approach handles the multidimensional SAGIN resources and learns the dynamic network characteristics. Simulation results confirmed the efficiency and convergence of the suggested technique.

As the heterogeneous multi-layer network requires advanced capacity-management techniques, Jiang and Zhu \cite{sagin5} suggested a low-complexity technique for computing the capacity among satellites and suggested a long-term optimal capacity assignment RL-based model to maximize the long-term utility of the system. 

By formulating the joint resources assignment problem as a joint optimization problem and using a DRL approach, Qiu et al. \cite{sagin6} proposed a software-defined satellite-terrestrial network to jointly manage caching, networking, and computing resources.

\subsection{Energy Managing}
\subsubsection{Definition \& limitations}

Recent advances in the connection between ground, aerial, and satellite networks such as SAGIN have increased the demand imposed on satellite communication networks. This growing attention towards satellites has led to increased energy consumption requirements. Satellite energy management thus represents a hot research topic for the further development of satellite communication. 

Compared with a GEO Satellite, an LEO satellite has restricted on-board resources and moves quickly. Further, an LEO satellite has a limited energy capacity owing to its small size \cite{energy1}; as billions of devices need to be served around the world \cite{energy2}, current satellite resource capability can no longer satisfy demand. To address this shortage of satellite communication resources, an efficient resource scheduling scheme to take full use of the limited resources, must be designed. As current resource allocation schemes have mostly been designed for GEO satellites, however, these schemes do not consider many LEO specific concerns, such as the constrained energy, movement attribute, or connection and transmission dynamics. 

\subsubsection{AI-based solutions}
Some researchers have thus turned to AI-based solutions for power saving. For example, Kothari et al. \cite{sat3} suggested the usage of DNN compression before data transmission to improve latency and save power. In the absence of solar light, satellites are battery energy dependent, which places a heavy load on the satellite battery and can shorten their lifetimes leading to increased costs for satellite communication networks. To optimize the power allocation in satellite to ground communication using LEO satellites and thus extend their battery life, Tsuchida et al. \cite{energy4} employed RL to share the workload of overworked satellites with near satellites with lower load. Similarly, implementing DRL for energy-efficient channel allocation in Satlot allowed for a 67.86\% reduction in energy consumption when compared with previous models \cite{energy5}. Mobile edge computing enhanced SatIoT networks contain diverse satellites and several satellite gateways that could be jointly optimized with coupled user association, offloading decisions computing, and communication resource allocation to minimize the latency and energy cost. In a recent example, a joint user-association and offloading decision with optimal resource allocation methodology based on DRL proposed by Cui et al. \cite{energy6} improved the long-term latency and energy costs.

\subsection{Other Applications}
\subsubsection{Handoff Optimization}

Link-layer handoff occurs when the change of one or more links is needed between the communication endpoints due to the dynamic connectivity patterns of LEO satellites. The management of handoff in LEO satellites varies remarkably from that of terrestrial networks, since handoffs happen more frequently due to the movement of satellites \cite{intro1}. Many researchers have thus focused on handoff management in LEO satellite networks. 

In general, user equipment (UE) periodically measures the strength of reference signals of different cells to ensure access to a strong cell, as the handoff decision depends on the signal strength or some other parameters. Moreover, the historical RSRP contains information to avoid unnecessary handoff. 

Thus, Zhang \cite{handoverconf} converted the handoff decision to a classification problem. Although the historical RSRP is a time series, a CNN was employed rather than an RNN because the feature map of historical RSRP has a strong local spatial correlation and the use of an RNN could lead to a series of wrong decisions, as one decision largely impacts future decisions. In the proposed AI-based method, the handoff was decreased by more than 25\% for more than 70\% of the UE, whereas the commonly used “strongest beam” method only reduced the average RSRP by 3\%. 

\subsubsection{Heat Source Layout Design}

The effective design of the heat sources used can enhance the thermal performance of the overall system, and has thus become a crucial aspect of several engineering areas, including integrated circuit design and satellite layout design. With the increasingly small size of components and higher power intensity, designing the heat-source layout has become a critical problem \cite{other0}. Conventionally, the optimal design is acquired by exploring the design space by repeatedly running the thermal simulation to compare the performance of each scheme \cite{other01,other02,other03}. To avoid the extremely large computational burden of traditional techniques, Sun et al. \cite{other1} employed an inverse design method in which the layout of heat sources is directly generated from a given expected thermal performance based on a DL model called Show, Attend, and Read \cite{other2}. Their developed model was capable of learning the underlying physics of the design problem and thus could efficiently forecast the design of heat sources under a given condition without any performing simulations. Other DL algorithms have been used in diverse design areas, such as mechanics \cite{other3}, optics \cite{other4}, fluids \cite{other5}, and materials \cite{other6}.

\subsubsection{Reflectarray analysis and design}

ML algorithms have been employed in the analysis and design of antennas \cite{reflect1}, including the analysis \cite{reflect2,reflect3} and design \cite{reflect4,reflect5} of reflectarrays. For example, NNs were used by Shan et al. \cite{reflect6} to forecast the phase-shift, whereas kriging was suggested to forecast the electromagnetic response of reflectarray components \cite{reflect7}. Support vector regression (SVR) has been used to accelerate the examination \cite{reflect8} and to directly optimize narrowband reflectarrays \cite{reflect9}. To hasten calculations without reducing their precision, Prado et al. \cite{reflect10} proposed a wideband SVR-based reflectarray design method, and demonstrated its ability to obtain wideband, dual-linear polarized, shaped-beam reflectarrays for direct broadcast satellite applications. 

\subsubsection{Carrier Signal Detection}

As each signal must be separated before classification, modulation, demodulation, decoding and other signal processing, localization, and detection of carrier signals in the frequency domain is a crucial problem in wireless communication. 

The algorithms used for carrier signal detection have been commonly based on threshold values and required human intervention \cite{carrier1,carrier2,carrier3,carrier4,carrier5,carrier6}, although several improvements have been made including the use of a double threshold \cite{carrier7,carrier8}. Kim et al. \cite{carrier12} proposed the use of a slope-tracing-based algorithm to separate the interval of signal elements based on signal properties such as amplitude, slope, deflection width, or distance between neighboring deflections. 

More recently, DL has been applied to carrier signal detection; for example, Morozov and Ovchinnikov \cite{carrier13} applied a fully connected NN for their detection in FSK signals, whereas Yuan et al. \cite{carrier14} used DL, to morse signals blind detection in wideband spectrum data. Huang er al. \cite{carrier15} employed a fully convolutional network (FCN) model to detect carrier signal in the broadband power spectrum. A FCN is a DL method for semantic image segmentation in which the broadband power spectrum is regarded as a 1D image and each subcarrier as the target object to transform the carrier detection problem on the broadband to a semantic 1D image segmentation problem \cite{carrier16,carrier17,carrier18}. Here, a 1D deep CNN FCN-based on was designed to categorize each point on a broadband power spectrum array into two categories (i.e., subcarrier or noise), and then position the subcarrier signals' location on the broadband power spectrum. After being trained and validated using a simulated and real satellite broadband power spectrum dataset, respectively, the proposed deep CNN successfully detected the subcarrier signal in the broadband power spectrum and achieved a higher accuracy than the slope tracing method.

\section*{Conclusion}
This review provided an overview of AI and its different sub-fields, including ML, DL, and RL. Some limitations to satellite communication were then presented and their proposed and potential AI-based solutions were discussed. The application of AI has shown great results in a wide variety of satellite communication aspects, including beam-hopping, AJ, network traffic forecasting, channel modeling, telemetry mining, ionospheric scintillation detecting, interference managing, remote sensing, behavior modeling, space-air-ground integrating, and energy managing. Future work should aim to apply AI, to achieve more efficient, secure, reliable, and high-quality communication systems.


%





\ifCLASSOPTIONcaptionsoff
  \newpage

\fi

\end{document}